\documentclass[floatfix,superscriptaddress,twocolumn,nobibnotes]{revtex4}
\usepackage{amsmath,amssymb,amsbsy,amsfonts,amsthm,verbatim}
\usepackage{graphicx,color}
\usepackage{bm}
\usepackage{booktabs,array}

\begin{document}
\title{Emergence of topological phases from the extension of two-dimensional lattice with nonsymmorphic symmetries}
\author{Pok-Man Chiu}
\affiliation{Department of Physics, National Tsing Hua University, Hsinchu City, Taiwan}
\affiliation{Institute of Physics, Academia Sinica, Taipei, Taiwan}
\author{Cheng-Yi Huang}
\affiliation{Institute of Physics, Academia Sinica, Taipei, Taiwan}
\author{Wan-Ju Li}
\affiliation{Institute of Physics, Academia Sinica, Taipei, Taiwan}
\author{Ting-Kuo Lee}
\affiliation{Institute of Physics, Academia Sinica, Taipei, Taiwan}
\date{\today}
\begin{abstract}
Young and Kane have given a great insight for 2D Dirac semimetals with nontrivial topology in the presence of nonsymmorphic crystalline symmetry. Based on one of 2D nonsymmorphic square lattice structures they proposed, we further construct a set of 3D minimal tight-binding models via vertically stacking the 2D nonsymmorphic lattice. Specifically, our model provides a platform to generate three topologically semimetallic phases such as Dirac nodal line semimetals, Weyl nodal line semimetals and Weyl semimetals. The off-centered mirror symmetry sufficiently protects nodal lines emerging within mirror-invariant plane with a nontrivial mirror invariant $n_{M\mathbb{Z}}$, whereas twofold screw rotational symmetry protects nontrivial Weyl nodal points with topological charge $C=2$. Interestingly, Weyl nodal loops are generated without mirror symmetry protection, where nontrivial ``drumhead" surface states emerge within loops. In the presence of both time-reversal and inversion symmetries, the emergence of weak topological insulator phases is discussed as well.
\end{abstract}

\maketitle

\section{Introduction}
Three dimensional topologically nontrivial semimetals (nodal-point and nodal line semimetals) have attracted great attention recently  \cite{Armitage18,Fang16,Kobayashi17}. Dirac and Weyl semimetals (DSM/WSM) contain isolated nodes with nontrivial topological properties in materials \cite{Armitage18}. For instance, nontrivial surface Fermi arc in WSMs connecting each pair of Weyl nodes with opposite topological monopoles has been discovered \cite{Hasan17}. In addition, the ultrahigh mobility \cite{Liang15}, giant magnetoresistance \cite{Liang15} and chiral anomaly effect \cite{Xiong15,Hasan17} in WSMs and DSMs have also been reported, all of which are of fundamental and applicational importance. As the second type of topological semimetals, topologically nontrivial band-crossing lines emerge (The Dirac nodal arc is also experimentally observed \cite{Wu16}) in the topological nodal line semimetals \cite{Fang16}. Materials hosting nodal lines have been experimentally confirmed \cite{Wu16,Neupane16,Bian16a,Hu16} and several intriguing properties are expected, including the drumhead surface states and special zero modes in quantum-oscillation measurements \cite{Fang16}. From these unique topological properties, an intuitive approach to achieve both types of nodal phases is highly desirable for material fabrications.

Conventional classifications of semimetals directly tackle the 3D space groups for searching candidates with topologically nontrivial semimetallic phases. WSMs and DSMs protected by symmorphic rotation \cite{Fang12a,Yang14} and reflection \cite{Chiu14} symmetries have been classified. Similar symmetry classifications are also performed for the nodal line semimetals \cite{Fang15,Chiu14}. Recently, the elementary excitations including nodal point, nodal line and even nodal-surface semimetals are exhaustively explored for all space groups with time-reversal symmetry (TRS) and spin-orbit coupling (SOC) \cite{Bradlyn16}. These influential works have provided useful clues for achieving distinct topological semimetallic phases. However, the complex nature of the 3D space groups makes it difficult for us to have an intuitive feeling about the essential  physics required for having these novel topological properties. Therefore, an alternative bottom-up method to generate 3D nontrivial phases by starting from minimal systems and keeping track of the evolution of topological properties could be an useful approach.

Recently, several works are devoted to generate different 3D topological phases by starting from a simple model and imposing distinct physical perturbations. First, (quasi) 2D space groups are extensively studied with crucial connections to 3D systems \cite{Wieder16,Kane15}. Then Behrends {\it et al.} \cite{Behrends17} introduced a periodically modulated potential to a 3D WSM to produce nodal-line semimetal phases. Most recently, Yang {\it et al.} considered a tetragonal lattice consisted of 2D square lattices stacked vertically \cite{Yang17}. Either by breaking the off-centered symmetry or the TRS, various topological insulating and semimetallic phases can be generated. Motivated by these works, we like to find out if those more sophisticated topological semimetals with nodal lines and loops could be also simply constructed by stacking a 2D model. Our strategy is to take advantage of  nontrivial topology in the 2D model. Hence we will use the minimal 2D nonsymmorphic lattice model with SOC developed by Young and Kane \cite{Kane15}. We shall only study models with TRS in this work. Various phases, such as Dirac-nodal-line (DNL) and Weyl-nodal-line (WNL) semimetals, WSM and weak topological insulator (WTI) are generated. Under specific conditions where only TRS is maintained, Weyl nodal loops are able to emerge, in contrast to those cases protected by mirror symmetries as well \cite{Bian16c}. Our bottom-up method may pave the way to engineer various topological semimetallic phases in materials. 

This paper is organized as follows. In Sec.~II, The minimal 2D nonsymmorphic lattice model is briefly introduced and its crucial physical features are discussed based on symmetry considerations. In Sec.~III, we construct several 3D tight-binding models by introducing different couplings of the 2D models along vertical directions. The topological phases of these models are analyzed. Discussions about the evolution from nodal points in 2D to nodal lines, nodal points in 3D, and the emergence of Weyl loops in the absence of off-centered mirror symmetry are provided in Sec.~IV. We conclude this work in  Sec.~V.

\section{\label{Kanemodel}2D Dirac semimetals with nonsymmorphic symmetry}
\begin{figure}
	\centering
		\includegraphics[width=245pt ]{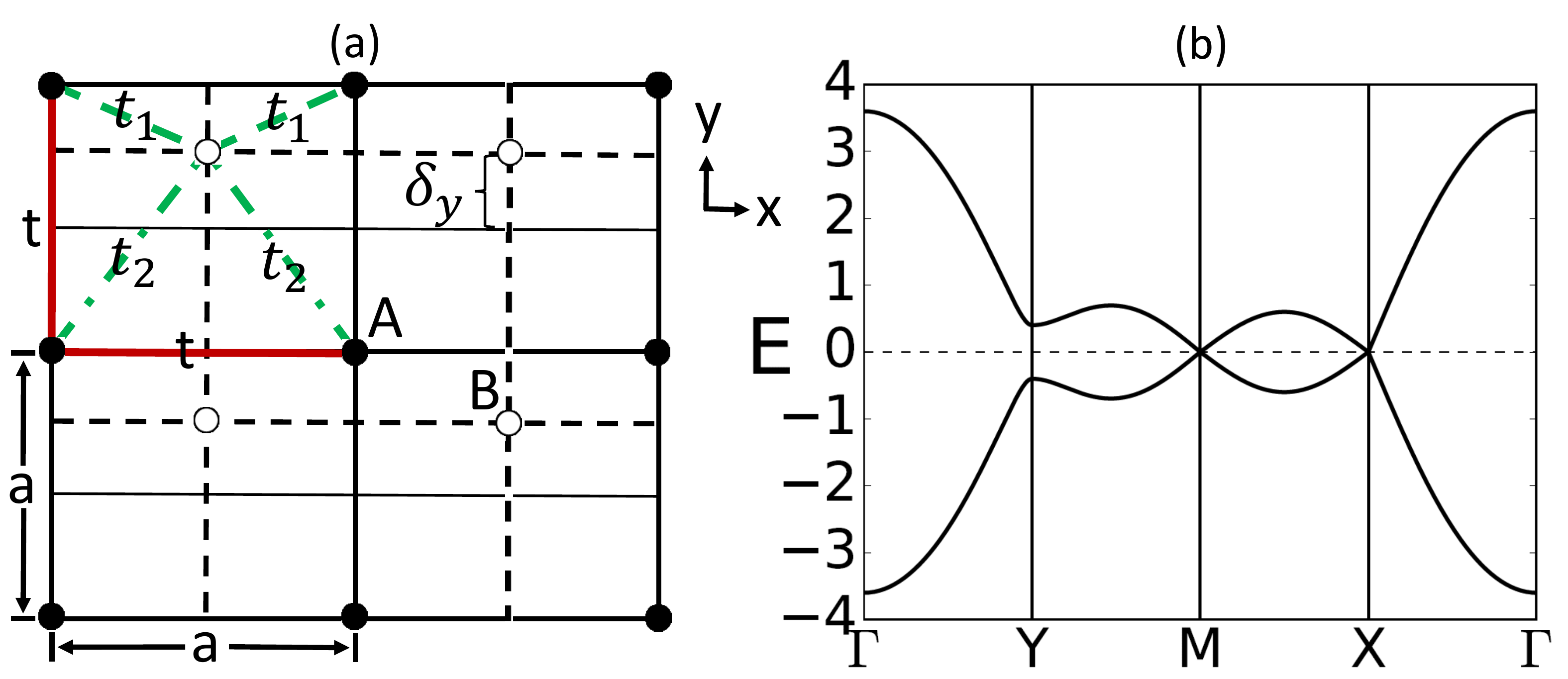}
	\caption{(Color online) (a) The schematic lattice structure. Solid (open) circle denotes A (B) sublattice. B atom is shifted a distance $\delta_y$ in $y$ axis away from center with a height $\delta_z$. (b) Band structure of $H_{2D}$. $\Gamma$, $Y$, $X$ and $M$ are at $(k_x,k_y)=(0,0)$, $(0,\pi)$, $(\pi,0)$ and $(\pi,\pi)$ respectively.}\label{band} 
\end{figure}

 We adopt one of the nonsymmorphic square lattice models proposed in Ref.~\cite{Kane15}. This model can be viewed as a minimal model for 2D Dirac semimetals with SOC in nonsymmorphic crystals. As shown in Fig. \ref{band}(a), we consider a square lattice consisted of A and B sublattices, where B atom is shifted a distance $\delta_y$ in $y$ axis away from the center of square lattice and it is also out of the plane with a height $\delta_z$. This lattice structure inherently has following symmetries: (1) inversion symmetry ($\mathcal{P}$), where the inversion center is at the middle of A and B atoms, (2) time-reversal symmetry ($\mathcal{T}$), (3) twofold screw rotation symmetry ($\mathcal{S}_{x}=\{C_{2x}|\dfrac{1}{2}0\}$), i.e., it is invariant to rotate by $\pi$ about $x$-axis and then translate a half lattice constant along $x$ direction, and (4) off-centered mirror symmetry ($\mathcal{M}^{\bot}_{x}=\mathcal{P}\mathcal{S}_{x}=\{M_x|\dfrac{1}{2}0\}$), which implies the mirror line passing through inversion center and requires a half translation along $x$ direction. Notably, $\mathcal{P}$, $\mathcal{S}_{x}$ and $\mathcal{M}^{\bot}_{x}$ form an important circular relation suggesting that it is necessary to break two of them simultaneously by a symmetry-breaking perturbation. According to the lattice geometry, in the presence of SOC, the four-band tight-binding model with the basis $(A\uparrow,A\downarrow,B\uparrow,B\downarrow)$ can be expressed as following: 
 \begin{align}
H_{2D}&=-2t(\cos k_{x}+\cos k_{y})
+\begin{bmatrix}
S&R&D&0\\
R^*&-S&0&D\\
D^*&0&-S&-R\\
0&D^*&-R^*&S
\end{bmatrix}\label{H2D}
\end{align}
where $D=-(t_{1}+t_{2}e^{ik_{y}})(1+e^{-ik_{x}})$, $S=2\lambda^{SO}\sin k_{x}$, $R=2\lambda_R(i\sin k_{x} +\sin k_{y} )$ and we set lattice constant $a=1$. $t_{1}$ ($t_{2}$) is AB sublattice hopping strength for short (long) bond. $t$ is a hopping strength between nearest neighbor atoms on the same sublattice. The intrinsic SOC, $\lambda^{SO}$, is inherently induced due to the lack of mirror symmetry about $y$ axis. $\lambda_R$ is the strength of Rashba coupling under broken mirror symmetry about $z$ axis similar to the case in silicenes \cite{Liu11}. Symmetry constrains for $H_{2D}$ are as following:
\begin{align}
\mathcal{T} &H_{2D}(k_x,k_y)\mathcal{T}^{-1}=H_{2D}(-k_x,-k_y),\notag\\
\mathcal{P}&H_{2D}(k_x,k_y)\mathcal{P}^{-1}=H_{2D}(-k_x,-k_y),\notag\\
\mathcal S_x(k_x)&H_{2D}(k_x,k_y)\mathcal S_x^{-1}(k_x)=H_{2D}(k_x,-k_y),\notag\\
\mathcal M^{\bot}_x(k_x) &H_{2D}(k_x,k_y)\mathcal M^{\bot}_x{}^{-1}(k_x)=H_{2D}(-k_x,k_y).\label{sym2d}
\end{align}
These symmetry operations are of the form:
\begin{gather}
\mathcal{T}=i\sigma_2K,\;\mathcal{P}=\begin{bmatrix}0&1\\1&0\end{bmatrix},\notag\\
\mathcal{S}_{x}(k_x)=\begin{bmatrix}0&1\\e^{ik_x}&0\end{bmatrix}\otimes i\sigma_1,\;
\mathcal{M}^{\bot}_{x}(k_x)=\begin{bmatrix}e^{ik_x}&0\\0&1\end{bmatrix}\otimes i\sigma_1,\label{sym}
\end{gather}
where $\bm{\sigma}$ are Pauli matrices acting on spin, explicit matrices act on AB sublattice and $K$ is complex conjugation. Since $t$ does not affect our result on topological properties, for simplicity we shall set it to be zero. The parameter set we use hereafter is: $t_1=1$, $t_{2}=0.8$, $\lambda_R=0.3$ and $\lambda^{SO}=0.1$. 

Notice that there are two gapless Dirac cones at $M (\pi,\pi)$ and $X(\pi,0)$, respectively [see Fig.~\ref{band}(b)]. In addition to TRS and inversion symmetry (IS), the interplay of screw rotation symmetry (SRS) in this nonsymmorphic lattice sufficiently guarantees the fourfold degeneracy of Dirac points at $M$ and $X$ in the presence of SOC. We can trace the number of occupied states with mirror eigenvalue $i$ in the lowest two bands along the mirror line $k_x=\pi$ and it changes by 2 across a Dirac point where the band inversion happens twice. The nontrivial mirror invariant \cite{Fang16,Chiu14},  $n_{M\mathbb{Z}}=N_i(k_1)-N_i(k_2)$, is hence $\pm2$, where $N_i(k)$ is number of occupied states with mirror eigenvalue $i$ and momenta $k_1$ and $k_2$ respectively locate at opposite sides of Dirac point. $n_{M\mathbb{Z}}$ implies that winding number around a Dirac cone is $\pm1$. In the following sections, this topological property is used to make the evolution from a 2D Dirac semimetal to a 3D topological semimetal. In particular, we study how the emergence of topological phases are influenced by breaking symmetry.

\section{Phases of Topological Semimetal in three dimension}

\begin{table*}
	\begin{ruledtabular}
		\begin{tabular}{ccccccc}
			$H'(k_x,\,k_z)$ & $\mathcal{T}$ & $\mathcal{P}$  & $\mathcal M^{\bot}_x$ & $\mathcal{S}_{x}$  & Phase & Topological invariant\\
			\hline
			$\sin k_{z} \Sigma_{31}$, $\cos k_{z}\Sigma_{00}$, $\cos k_{z}\Sigma_{10}$ & + & + & + & + & DNL & $M\mathbb{Z}$\\
			\hline
			$\sin k_{z}\Sigma_{01}$, $\sin k_{z}\Sigma_{11}$, $\sin k_{z}\Sigma_{22}$, $\sin k_{z}\Sigma_{23}$, $\cos k_{z}\Sigma_{30}$& + & - & + & - & WNL & $M\mathbb{Z}$\\
			\hline
			$\sin k_{z}\Sigma_{02}$, $\sin k_{z}\Sigma_{03}$, $\sin k_{z}\Sigma_{12}$, $\sin k_{z}\Sigma_{13}$, $\sin k_{z}\Sigma_{21}$ & + & - & - & + & WSM & $\mathbb{Z}$\\
			\hline
			$\cos k_z\Sigma_{20}$, $\sin k_{z}\Sigma_{32}$, $\sin k_{z}\Sigma_{33}$ & + & + & - & - & WTI/NSM & $\mathbb{Z}_2$
		\end{tabular}
	\end{ruledtabular}
	\caption[]{\label{T1} Summary of possible $H'(k_x,\,k_z)$ with time-reversal invariance. Symmetry analysis is based on Eq.~\ref{symarg} and $\Sigma_{ij}=\tau_i\sigma_j$. $\sigma_i$ are Pauli matrices for $i=0\sim3$ acting on spin. $\tau_i$ acting on AB sublattice are defined by $\tau_0=I_{2\times2}$, $\tau_1=\begin{bmatrix}0&1+e^{-ik_x}\\1+e^{ik_x}&0\end{bmatrix}$ , $\tau_2=\begin{bmatrix}0&1-e^{-ik_x}\\1-e^{ik_x}&0\end{bmatrix}$ and $\tau_3=\begin{bmatrix}1&0\\0&-1\end{bmatrix}$. The symbol $+(-)$ indicates that the symmetry is preserved (broken). DNL (WNL) denotes Dirac (Weyl) nodal line semimetal phase. WSM (WTI, NSM) denotes Weyl semimetal (weak topological insulator, normal semimetal) phase. $M\mathbb{Z}$ denotes $\mathbb{Z}$ index defined in mirror plane.}
\end{table*}

In order to investigate the topological phases of 3D topological semimetals in our minimal model, we simply construct a layered structure by vertically stacking the 2D lattice in Sec.~\ref{Kanemodel}. Consequently, the resulting 3D lattice structure holds the similar symmetry operations in 2D structure and the expressions of symmetry operations in Eq.~(\ref{sym}) are valid as well. An interlayer coupling $H'(k_x,\,k_z)$ is added to the $H_{2D}$ to have $H_{3D}(k_x,k_y,k_z)=H_{2D}(k_x,k_y)+H'(k_x,\,k_z)$. Similar to Eq.~\ref{sym2d}, we have the symmetry conditions for $H_{3D}$ :
\begin{align}
\mathcal{T}&H_{3D}(k_x,k_y,k_z)\mathcal{T}^{-1}=H_{3D}(-k_x,-k_y,-k_z),\notag\\
\mathcal{P}&H_{3D}(k_x,k_y,k_z)\mathcal{P}^{-1}=H_{3D}(-k_x,-k_y,-k_z),\notag\\
\mathcal S_x(k_x)&H_{3D}(k_x,k_y,k_z)\mathcal S_x^{-1}(k_x)=H_{3D}(k_x,-k_y,-k_z),\notag\\
\mathcal M^{\bot}_x(k_x) &H_{3D}(k_x,k_y,k_z)\mathcal M^{\bot}_x{}^{-1}(k_x)=H_{3D}(-k_x,k_y,k_z).\label{symarg}
\end{align}
Notice that the dependence of $k_x$ in $H'(k_x,\,k_z)$ is necessary for interlayer AB sublattice couplings in order to globally preserve or break $\mathcal S_x$ or $\mathcal M^{\bot}_x$ in momentum space. We focus on the $k_x=\pi$ plane since nonsymmorphic symmetry plays a crucial role to protect gapless features in the system. Our conclusion would not be changed while keeping a fully gap elsewhere $k_x\neq \pi$.

We systematically investigate the topological phases via preserving or breaking nonsymmorphic crystalline symmetries. In our 3D model, topological indices $n_{M\mathbb{Z}}$ and $n_{\mathbb{Z}}$ \cite{Fang16,Chiu14} are useful to investigate the topology of nodal lines and nodal points in the presence and absence of off-centered mirror symmetry, respectively. Table~\ref{T1} summarizes possible $H'(k_x,\,k_z)$ with preserving TRS. When $H'(k_x,\,k_z)$ breaks certain nonsymmorphic crystalline symmetries, for example via crystal distortion, various forms of stacking and strains, it will generate different topological phases. Below we will consider three topological semimetal phases, i.e., DNL semimetal, WNL semimetal and WSM, and WTI phases.

\subsection{Dirac nodal line semimetals}

\begin{figure}
	\centering
		\includegraphics[width=245pt]{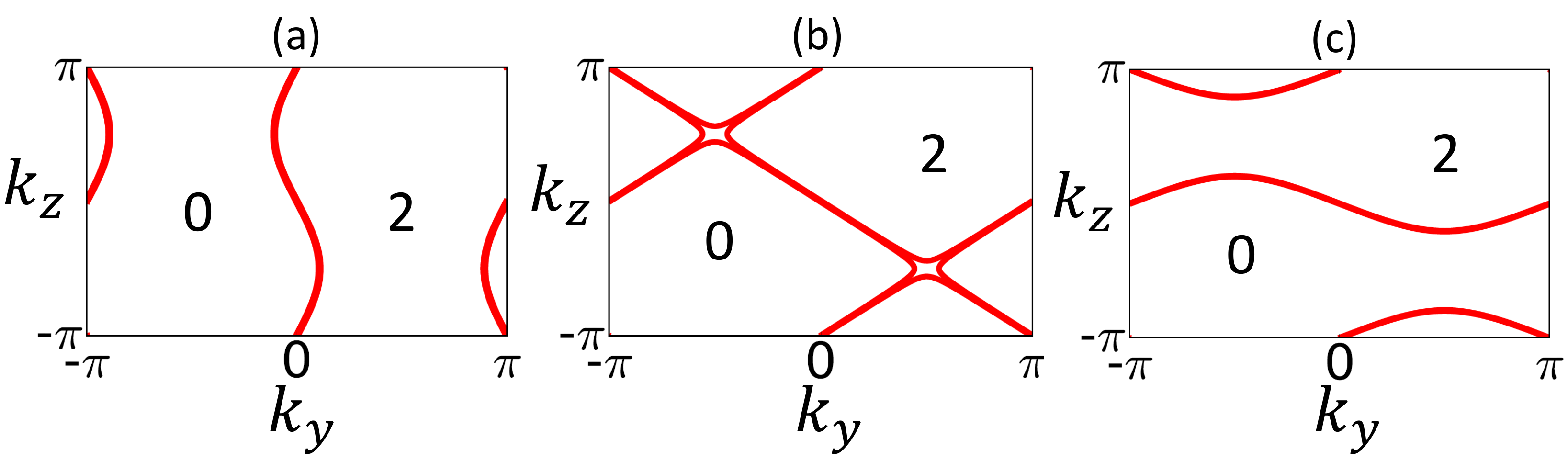}
	\caption{\label{DNL} Evolutions of Dirac nodal lines within $k_{x}=\pi$ plane where $H'(k_x,\,k_z)=2t_{z}\sin k_{z}\Sigma_{31}$. (a), (b) and (c) show that nodal lines change shape when $t_z=$ 0.1, 0.3 and 0.5, respectively. (b) shows that the nodal lines touch each other at certain momenta incidentally, where a Lifshitz transition occurs. The inserted integers denote the number of occupied states with mirror eigenvalue $i$.} 
\end{figure}

The phase of DNL semimetal emerges with fourfold degenerate nodal lines, which is also known as double nodal lines \cite{Yang17,Fang15,Fang16}, when the four symmetries of Eq.~\ref{symarg} are preserved. In contrast to the DNL semimetals without SOC, the presence of SOC requires additional crystalline symmetries to protest two doubly degenerate band crossing \cite{Fang15,Fang16}. The necessary symmetries can be glide \cite{YChen16}, twofold screw rotation \cite{Fang15,Fang16,Liang16} or off-centered mirror symmetry \cite{Yang17}. In this work, DNLs are mainly supported by the interplay of off-centered mirror symmetry, IS and TRS. Hence DNLs are robust on mirror-symmetry invariant plane $k_x=\pi$. In the second row of Tabel~\ref{T1}, $\sin k_{z} \Sigma_{31}$ indicates a Rashba-like coupling along $z$ direction, $\cos k_z\Sigma_{10}$ is for an interlayer AB hopping term, while $\cos k_{z}\Sigma_{00}$ for a simple vertical hopping term. Take $H'(k_x,\,k_z)=2t_z\sin k_{z} \Sigma_{31}$ for example, in $k_x=\pi$ plane, $H_{3D}$ simply becomes $H_{3D}=2\tau_3\sigma_1(\lambda_R\sin k_y+t_z\sin k_z)$ and energy dispersions are $\pm2|\lambda_R\sin k_y+t_z\sin k_z|$, where gapless nodal lines with fourfold degeneracy lie on momenta satisfying $\lambda_R\sin k_y=-t_z\sin k_z$. Nodal lines confined within the mirror plane and linking two of time-reversal invariant momenta can change their shape. As shown in Fig.~\ref{DNL}, these gapless nodal lines deform with varying $t_z$. Interestingly, a Lifshitz transition, where each line change its connected time-reversal invariant momenta, could take place when $t_z=\pm\lambda_R$. We can count the number of occupied states with mirror eigenvalue $i$ within this plane to trace how many times band inversion happens. Fig.~\ref{DNL} shows that any crossing of a nodal line will change the occupied states by 2, which indicates $n_{M\mathbb{Z}}=\pm2$. The nontrivial $n_{M\mathbb{Z}}$ in the original 2D model provides the topological property of DNL phase. As a consequence of the preservation of crystalline symmetry, topology in 2D model naturally leads to the evolution of 3D DNL.

\subsection{Weyl nodal line semimetal}

\begin{figure}
	\centering
		\includegraphics[width=245pt ]{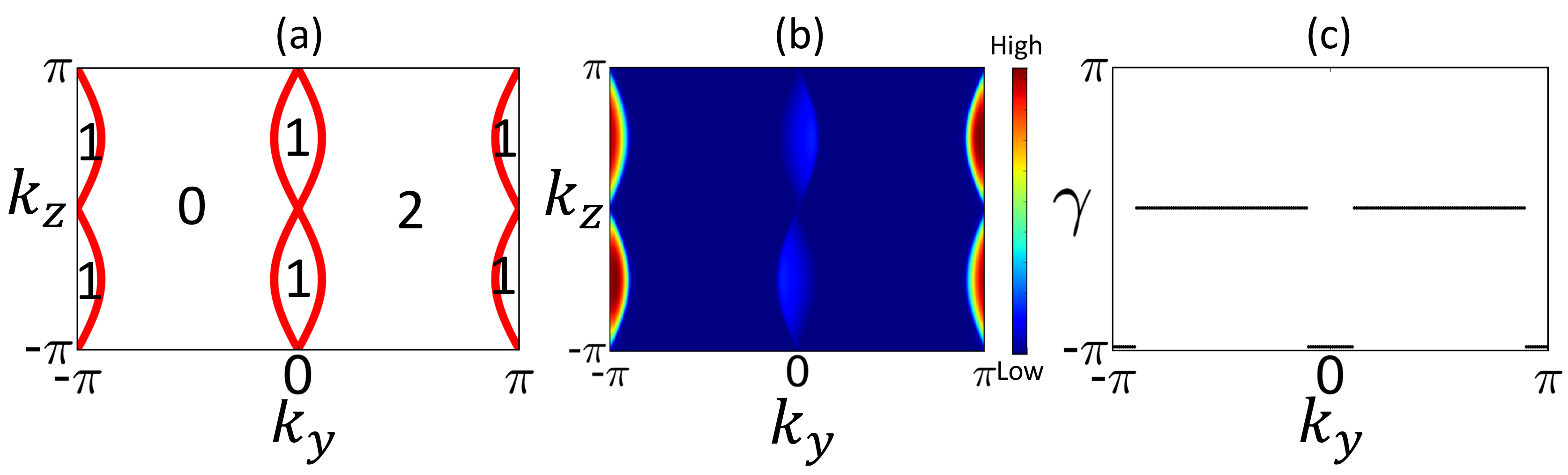}
	\caption{\label{WNLfig}(Color online) Weyl nodal phase where $H'(k_x,\,k_z)=2t_z\sin k_z\Sigma_{01}$ and $t_{z}=0.1$. (a) Weyl nodal lines within $k_x=\pi$ plane. The inserted integers denote the number of occupied states with mirror eigenvalue $i$. (b) Fermi surface of surface states on (100) surface. Color indicates the intensity of spectral weight. (c) Berry phase $\mathcal{\gamma}(k_y,k_z=\pi/2)$ as a function of $k_y$. A $\pm\pi$ jump indicates the existence of nontrivial surface states.} 
\end{figure}
It is quite surprising to find the nodal lines even when both $\mathcal P$ and $\mathcal S_x$ are broken as shown in the third row of Table~\ref{T1}. In contrast to the case in Sec. III-A, nodal lines within $k_x=\pi$ plane becomes doubly degenerate and surface states would emerge between nodal lines. We called these nodal lines in this section ``Weyl nodal lines" (or, in general, so-called topological nodal lines \cite{Burkov11,Chiu14,Fang16,Yu17}) in order to distinguish from DNLs in Sec. III-A.

In the absence of $\mathcal P$ and $\mathcal S_x$, each DNL splits into two WNLs, $\mathcal T$ and $\mathcal{M}^{\bot}_{x}$ are sufficient to support WNLs. Table~\ref{T1} shows five terms corresponding to this category. Take $H'(k_x,\,k_z)=2t_z\sin k_z\Sigma_{01}$ for instance, $H_{3D}$ within $k_x=\pi$ plane becomes $H_{3D}=2\lambda_R\tau_3\sigma_1\sin k_y+2t_z\tau_0\sigma_1\sin k_z$ and energy dispersions are $E=\pm2|\lambda_R\sin k_y\pm t_z\sin k_z|$. There are four doubly degenerate nodal lines satisfying $\lambda_R\sin k_y=\pm t_z\sin k_z$ in this plane [see Fig.~\ref{WNLfig}(a)]. Surface states on (100) surface especially emerge in the region between nodal lines that has an odd number of occupied states with mirror eigenvalue $i$ [see Fig.~\ref{WNLfig}(a)-(b)]. The emergence of non-trivial surface states can be confirmed by verifying Berry phase and $n_{M\mathbb Z}$. Evolution of Berry phase $\gamma(k_y,k_z)$ \cite{Berry84,Zak89,Kariyado13,Xiao10,Yu11} integrated over $k_x$ (see Appendix A. for more details) is showed in Fig.~\ref{WNLfig}(c). $\gamma(k_y,k_z=\pi/2)$ has a dramatic $-\pi$ jump with varying $k_y$, which indicates the existence of non-trivial surface states. Meanwhile, $n_{M\mathbb{Z}}=\pm1$ across a nodal line, where band inversion obviously happens once and the nodal line is a boundary of different topological regions. In general, WNLs could accidentally cross each other somewhere. Introducing a symmetry-preserving extra mass term, e.g., a staggered potential $V\Sigma_{30}$ on AB sublattice, can remove this accidental degeneracy. WNLs are, however, protected by off-centered mirror symmetry in spite of the strength of $V$.

\subsection{Weyl semimetals}
\begin{figure}
	\centering
		\includegraphics[width=245pt]{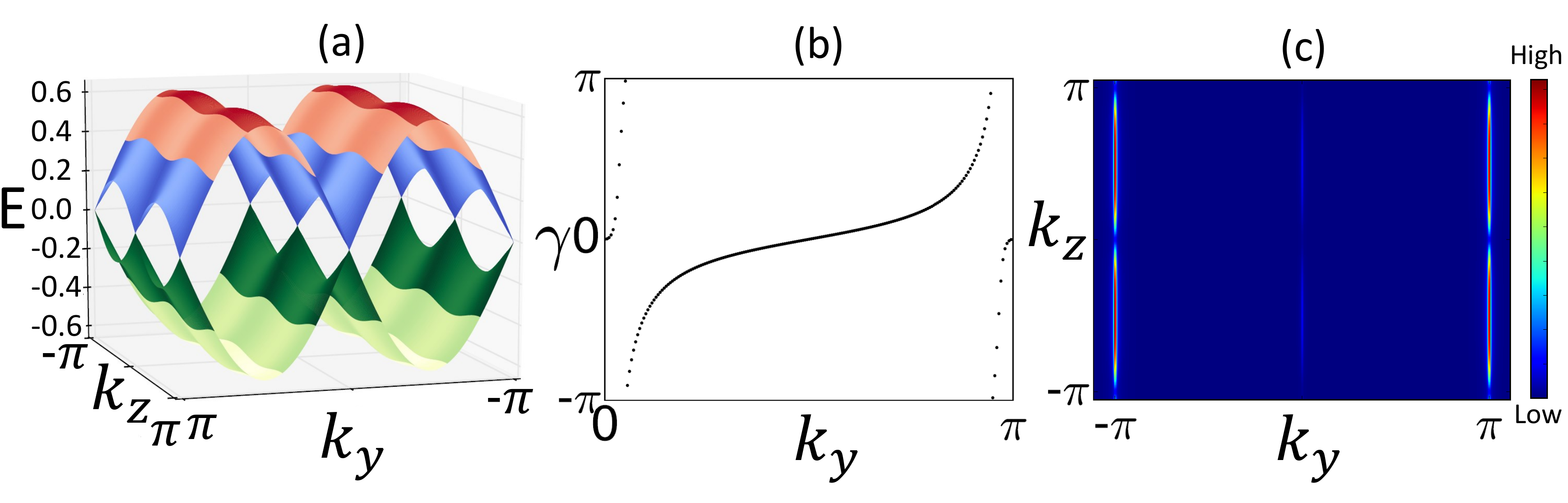}
	\caption{\label{WSMband}(Color online)(a) 3D band structure on the $k_x=\pi$ plane, where $H'(k_x,\,k_z)=2t_{z}\sin k_{z}\Sigma_{03}$ and $t_z=0.1$. There are four Weyl nodes within this plane. (b) The flow of Berry phase $\gamma$ around Weyl node at $(k_y,k_z)=(0,0)$ indicates topological charge $C=2$. (c) Fermi surface when opening (100) surface. Four Fermi arcs connect four time-reversal invariant momenta. Color denotes intensity of spectral weight.} 
\end{figure}

In this section, we are going to investigate the topological WSM phase in the presence of both TRS and SRS. It is known that Weyl nodes could emerge once IS or TRS is broken \cite{Burkov11a,Burkov11b}. Nodal lines lose the protection of off-centered mirror symmetry within $k_x=\pi$ plane and are fully gapped except at time-reversal invariant momenta. Weyl nodes in our case locate at time-reversal invariant momenta $\Lambda_{k_x=\pi}$ within $k_x=\pi$ plane, due to the protection of both TRS and SRS. Large surface Fermi arcs only survive on (100) surface, sharply contrasting with ``Kramers Weyl fermions", where Weyl nodes locate at all time-reversal invariant momenta in whole 3D Brillouin zone \cite{Chang16b}, which harbors large Fermi arcs on any surface plane. It is worth to point out that, thanks to twofold SRS, a combined symmetry operation $\tilde{\mathcal S}(k_x)=\mathcal S_x(k_x)\mathcal T$ provides a local ``Kramers-like" doubly degeneracy within $k_x=\pi$ plane, where $[\tilde{\mathcal S}(k_x=\pi)]^2=-1$ \cite{Wieder16,Chang17prl}. The fourfold degenerate Weyl nodes, therefore, are crystalline-symmetry-protected. In the fourth row of Table I, there are five terms classified in WSM phase. Notice that these terms have certain strong spin-orbit coupling in order to satisfy SRS. To explicitly demonstrate the physical property of Weyl nodes, take $H'(k_x,\,k_z)=2t_z\sin k_z\Sigma_{03}$ for example. Fig.~\ref{WSMband}(a) shows the energy dispersion within $k_x=\pi$ plane, where four Weyl nodes reside. Topological charge or Chern number is $\pm2$ for each Weyl node similar to double-Weyl fermions \cite{Tang17}. In Fig.~\ref{WSMband}(b), topological charge is determined by means of Wilson loop \cite{Yu11,Soluyanov15,Gresch17,Bouhon17a,Bouhon17b} on a closure sphere enclosing a Weyl node. Chern number $C$ can be identified by the total flow of Berry phase $\gamma$ (divided by $2\pi$) integral of a circular loop $\mathcal L$ on a sphere with varying a polar angle $\theta$ from $0$ to $\pi$, i.e., $2\pi C = \Delta\gamma[\mathcal L(0\rightarrow\pi)]$, where $\gamma$ is Berry phase calculated by occupied states. The detail numerical method is shown in Appendix A. In order to understand the topology associated with the Weyl node, we examine $H_{3D}$ around $(k_x,k_y,k_z)=(\pi,0,0)$ and take a ${\bm k}\cdot {\bm p}$ expansion up to $k$ linear terms. Here we can set $\lambda^{SO}=0$ as it will not affect our discussion below. After a suitable unitary transformation, $H_{3D}$ can be further separated into two subspaces,
\begin{align}
H_{\eta}=&-(t_1+t_2)\alpha_3k_x+2\lambda_R(\eta\alpha_2k_x+\alpha_1k_y)-2\eta t_z\alpha_3k_z\notag\\
=&\bm{ d}_{\eta}\cdot \bm{\alpha},
\end{align}
where $\bm{\alpha}$ are Pauli matrices, ${\bm d}_{\eta}=(2\lambda_Rk_y\,,\,2\eta\lambda_Rk_x\,,\,-(t_1+t_2)k_x-2\eta t_zk_z)$ and $\eta=\pm1$. Both $H_{\eta}$ precisely describe two spin-1/2 Weyl fermions with the topological nature of Chern number 1. As a result, total Chern number is 2 for this double Weyl node. A Weyl node carrying Chern number 2 suggests the emergence of 2 surface Fermi arcs connecting another node with an opposite topological charge. Fig.~\ref{WSMband}(c) shows Fermi arcs connecting $(k_y,k_z)=(0,0)$ with $(0,\pi)$ or $(\pi,0)$ with $(\pi,\pi)$ in opposite $z$ directions.

\subsection{Weak topological insulators}

\begin{figure}
	\centering
		\includegraphics[width=245pt ]{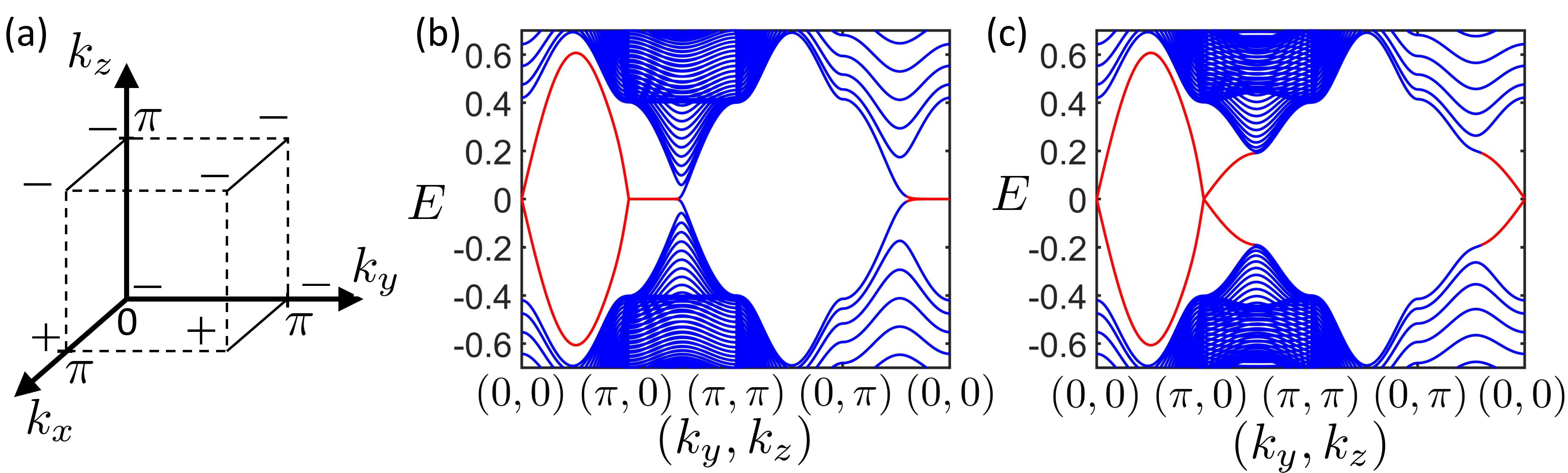}
	\caption{\label{TSMband}(Color online) (a) Schematic diagram depicting $\delta_n$ at each time-reversal invariant momenta $\Lambda_n$, where $H'(k_x,\,k_z)=t_{z}\cos k_{z}\Sigma_{20}+2t_{z}'\sin k_{z}\Sigma_{33}$ and $t_z=0.2$. Symbol $+(-)$ denotes $\delta_n=1(-1)$. Energy dispersions of slab structure opening (100) surface are shown when $t_z'=$ (b) 0 and (c) 0.1. Red (blue) curves denote bands of surface (bulk) states. }
\end{figure}

In this last section, we are going to discuss the topological phases in the presence of both TRS and IS. When both nonsymmorphic symmetries $\mathcal{M}^{\bot}_{x}$ and $\mathcal S_x$ are broken, the symmetry class reduces to symmorphic symmetry. There are accidental gapless nodes residing within $k_x=\pi$ plane although and $\mathbb Z_2$ index is rather appropriate to identify the topology than $\mathbb Z$ index or Chern number, suggested by Yang and Nagaosa \cite{Yang14}. However, both $\mathcal T$ and $\mathcal P$ do not sufficiently support the robustness of gapless feature and a gap opening is allowed. As shown in the last row of Table I, there are three possible $H'(k_x,\,k_z)$ terms respecting both TRS and IS. For the first case, $H'(k_x,\,k_z)=t_z\cos k_z\Sigma_{20}$, there are gapless nodal points away from time-reversal invariant momenta. $\mathbb Z_{2}$ invariant is well-defined at fully gapped time-reversal invariant momenta $\Lambda_{n}$. We preform the three-dimensional $\mathbb Z_{2}$ invariant, $(\nu_0;\nu_1\nu_2\nu_3)$, to determine the topology of system. Strong index $\nu_0$ by definition is
\begin{align}
(-1)^{\nu_0}=\prod_{\forall\Lambda_{n}}\delta_n,
\end{align} 
and weak indices $\nu_i$, $i=1,2,3$, are expressed as
\begin{align}
(-1)^{\nu_i}=\prod_{\Lambda_{n}\in k_i=\pi}\delta_n,
\end{align}
where $\delta_n=\prod_{m}\xi_{m}(\Lambda_n)$ defined at the $n$-th time-reversal invariant momenta $\Lambda_n$ and $\xi_m$ is the $m$-th parity eigenvalue belonging to two occupied states. We find then
\begin{align}
(-1)^{\nu_0}&=(-1)^{\nu_1}=1,\,(-1)^{\nu_2}=-1,\notag\\
(-1)^{\nu_3}&=\text{sgn}(t_1^2-t_2^2).
\end{align}
As a consequence, we can determine the different parameter regimes for $H_{3D}$. When $t_1>t_2$, $(\nu_0;\nu_1\nu_2\nu_3)=(0;010)$; when $t_1<t_2$, $(\nu_0;\nu_1\nu_2\nu_3)=(0;011)$. These two regimes are ``weak topological insulators'', where those gapless points are unstable. For example, if $H'(k_x,\,k_z)=t_z\cos k_z\Sigma_{20}+2t_z'\sin k_z\Sigma_{33}$, it will lead to gap-opening immediately once $t_z'\neq0$, where $\mathbb Z_2$ indices do not change. Surface states only emerge on certain surface plane implied by weak $\mathbb Z_2$ indices. Fig.~\ref{TSMband}(a) showing $\delta_n$ at each $\Lambda_n$ implies there could be surface states emerging on (100) or (001) surface and Fig.~\ref{TSMband}(b) shows energy dispersion in slab structure opening (100) surface as $t_1>t_2$. Notably, two pieces of nodal-line-like surface states emerge along $k_y=0$ and $\pi$ lines. These surface states have linear dispersion along $k_y$ and only can propagate in $y$ direction with a high mobility on surface. However, when an extra symmetry-allowed coupling exists, e.g., $\sin k_z\Sigma_{33}$, the nodal-line-like surface states become two surface Dirac cones at $(k_y,k_z)=(0,0)$ and $(\pi,0)$ simultaneously [see Fig.~\ref{TSMband}(c)]. This is consistent with the suggestion of weak $\mathbb Z_2$ indices.

\section{Discussion}
The symmetry condition in the layered structure we constructed in this work is, however, less complicated than the cases in Ref. \cite{Yang17}. Our minimal 3D tight-binding model can generate multiple topological phases for nodal semimetals with or without nonsymmorphic symmetries and gives an insight of understanding the emergence of nodal lines and nodal points in materials. This model provides a platform to further study novel excitations hidden in topological nodal semimetals.

It has been known that the stabilization of Dirac points in 2D and 3D system must be held by crystalline symmetry protection \cite{Yang14,Gao16}. Here we show that the two Dirac nodal points at $k_x=\pi$ in the 2D model can generate DNLs by vertically stacking layers as long as nonsymmorphic symmetry is preserved, even if there is interlayer strong SOC, e.g., $\sin k_z \Sigma_{31}$ in Table.~\ref{T1}. The mechanism of DNLs in our work contrasts to the case proposed in topological/normal insulator superlayer structure \cite{Burkov11b}, where nodal lines are obtained via the modulation between trivial and nontrivial topological regions. Our method provides another approach to realize Dirac nodal semimetals for layered materials.

\begin{figure}
	\centering
			\includegraphics[width=8.5cm]{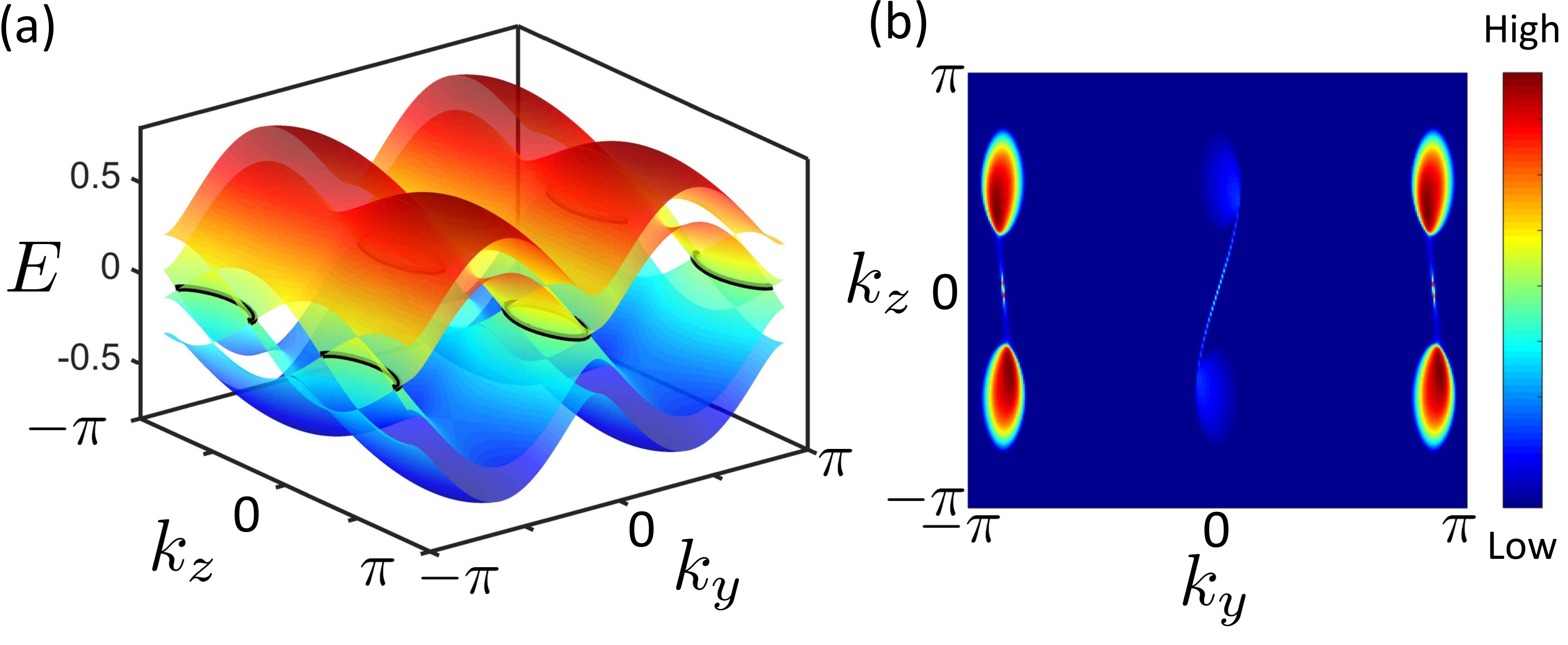}
	\caption{\label{loops}(Color online) (a) 3D band structure within $k_x=\pi$ plane, where $H'(k_x,\,k_z)=2t_{z}\sin k_{z}\Sigma_{01}+t_z'\cos k_z\Sigma_{20}$ and $t_z=t_z'=0.1$. Black loops indicate Weyl loops with the same energy. Weyl nodes reside away from zero energy level. (b) Fermi surface of surface states on (100) surface. ``Drumhead" surface states emerge in Weyl loops and Fermi arcs connect two loops. Color denotes the intensity of spectral weight.}
\end{figure}

The phenomenon of DNLs splitting into WNLs by breaking both IS and SRS can be understood that gap-opening between DNLs is caused by broken-IS while off-centered mirror symmetry protects the robustness of WNLs. Once the off-centered mirror symmetry is slightly broken in the presence of a weak perturbation, e.g., strain, the WNLs could become Weyl nodal loops. Take $H'(k_x,\,k_z)=2t_{z}\sin k_{z}\Sigma_{01}$ in WNL phase in Table I for example. If we introduce an extra coupling, e.g., $t_z'\cos k_z\Sigma_{20}$, into $H'$ to break off-centered mirror symmetry simultaneously, Weyl nodal loops are shown in Fig.~\ref{loops}(a) with $t_z=t_z'=0.1$. Interestingly, as shown in Fig.~\ref{loops}(a), nodal loops with zero energy emerge within $k_x=\pi$ plane and, meanwhile, Weyl nodes with Chern number $\pm1$ reside away from zero energy level. Nodal loops viewed as the intersection of two Weyl nodes with an opposite topological charge enclose topologically nontrivial surface states [see Fig.~\ref{loops}(b)], similar to so-called ``drumhead" surface states in TlTaSe$_2$ \cite{Bian16c} and Ca$_3$P$_2$ \cite{Chan16}. Notably, this is an interesting situation where nodal loops and nodal points with nonzero topological charge coexist simultaneously in a topological system. There emerge nontrivial surface Fermi arcs connecting loops as well. Surprisingly, these Weyl loops are robust even under large $t_z'$ in this case. We emphasize here, in general, the emergence of nontrivial nodal loops {\it does not} necessarily require the existence of mirror symmetry.

The 2D lattice model in this work is case II discussed by Young and Kane \cite{Kane15}. This 2D Dirac semimetal with SOC shares the same space group as the Bi(110) monolayer \cite{Bian14,Koroteev08,Bian17}, which might be experimentally realized in future. Many terms in Table I correspond to specific hopping mechanisms. For instance, the $\cos k_z\Sigma_{10}$ for the DNL phase can be realized by interlayer inter-sublattice hopping while the $\cos k_z\Sigma_{30}$ for the WNL phase corresponds to the interlayer intra-sublattice hopping where orbitals in two sublattices have different symmetries. Accordingly, distinct topological phases can be constructed with the guidance from Table I, which will help to engineer 3D topological materials.

\section{Conclusion} 
In conclusion, we theoretically provide a minimal tight-binding model of layered structure, which can straightforward generate 3D topological semimetal phases with interplay of nonsymmorphic crystalline symmetries. Our 3D models can demonstrate topologies of Dirac nodal line semimetals, Weyl nodal line semimetals, Weyl semimetals and weak topological insulators. In the presence of off-centered mirror symmetry, nodal lines emerge within mirror-invariant plane with nontrivial winding number. On the contrary, in the absence of off-centered mirror symmetry, nodal lines are gapped and nontrivial nodal points could survive via the protection of screw rotational symmetry with double topological charge, $C=2$. Surprisingly, Weyl nodal loops are present even only TRS is preserved without mirror-symmetry protection. WTIs generated in the presence of both TRS and IS are discussed, where nodal-line-like surface states or even surface Dirac cones emerge.

\acknowledgements
The authors like to thank Dr. P. J. Chen for valuable discussions.  The work is supported in part by  Taiwan
Ministry of Science and Technology Grant 106-2119-M-001-028. Part of calculation was supported by the National Center for High Performance Computing in Taiwan.

\appendix
\section{Numerical method for Wilson loops and Chern number}
We give a brief review for the numerical Wilson loop method \cite{Yu11,Soluyanov15,Gresch17,Bouhon17a,Bouhon17b} to calculate both Berry phase and Chern number.  Berry phase $\gamma$ is calculated by the integration over $[-\pi,\pi]$ along a path in momentum space and Chern number is determined by the evolution of Berry phase $\gamma$ with varying integral loop on a spherical surface enclosing a targeted nodal point \cite{Weng15}.

Firstly, we define the Berry-Wilczek-Zee connection \cite{Berry84,Wilczek84} using the cell periodic Bloch occupied eigenstates $\lvert u_{n,\boldsymbol{k}} \rangle$ with $n=1,...,N$, and $N$ is the number of occupied band, i.e., $\mathcal{A}_{mn,\mu}=\langle u_{m,\boldsymbol{k}} \lvert \partial_{k_{\mu}}\lvert u_{n,\boldsymbol{k}}\rangle$, where $\mu=x,y,z$. In both calculating Berry phase and Chern number, we employ the following useful relation between Wilson loop and Berry phase 
\begin{align}
e^{i\gamma[\mathcal{L}]} &= P\exp\bigg[-\int_{\mathcal{L}}d\boldsymbol{k}\cdot \mathrm{Tr} \mathcal{A}\bigg]
= \det \mathcal{W}[\mathcal{L}].
\end{align}
where $\mathcal{W}[\mathcal{L}]=P\exp[-\int_{\mathcal{L}}d\boldsymbol{k}\cdot \mathcal{A}]$ is the Wilson loop, which can be calculated numerically. Here $P$ means the integration order is counter-clockwise. $\mathcal{L}$ denotes the integration path. 

In the method of Wilson loop, $\mathcal{W}[\mathcal{L}]$ can be expressed as discretized product form $\mathcal{W}_{mn}[\mathcal{L}]=\langle u_{m,{\bf k}} \lvert \mathcal{W}[\mathcal{L}]\lvert u_{n,{\bf k}}\rangle$, where ${\bf k}$ is a momentum on $\mathcal{L}$. The Wilson loop operator is hence defined as $\mathcal{W}[\mathcal{L}]=P\prod_{\boldsymbol{k}\in \mathcal{L}}\mathcal{P}(\boldsymbol{k})$, where the projector operator is $\mathcal{P}(\boldsymbol{k})=\sum^{N}_{m,n}\lvert u_{n,\boldsymbol{k}} \rangle \langle u_{m,\boldsymbol{k}} \lvert$. For Chern number calculation, $\mathcal{L}(\theta)$ denotes a circular path on a spherical surface at a polar angle $\theta$, where $\theta=0$ means along $k_{z}$ axis \cite{Bouhon17a,Bouhon17b}. The Chern number is determined by the total flow of Berry phase $\gamma$ from $\theta=0$ to $\pi$, i.e. $2\pi C=\Delta\gamma[\mathcal{L}(0\rightarrow\pi)]$.

\bibliography{reference}
\bibliographystyle{prsty}

\end{document}